\documentclass{PoS}
\usepackage{amsmath}
\usepackage{amssymb}
\usepackage{graphicx}
\usepackage{epsfig}
\usepackage{latexsym,bm}
\usepackage{color}
\usepackage{multirow}%
\usepackage{cite}

\newcommand{\be}{\begin{equation}} \newcommand{\ee}{\end{equation}}
\newcommand{\ba}{\left(\begin{array}{c}}
\newcommand{\ea}{\end{array}\right)}
\newcommand{\bea}{\begin{eqnarray}} \newcommand{\eea}{\end{eqnarray}}

\newcommand{\al}{&\!\!\!\!}

\newcommand{\bma}{\left(\begin{matrix}}
\newcommand{\ema}{\end{matrix}\right)}
\newcommand{\bqa}{\begin{eqnarray}}
\newcommand{\eqa}{\end{eqnarray}}
\newcommand{\bqaa}{\begin{eqnarray*}}
\newcommand{\eqaa}{\end{eqnarray*}}

\title{Resonance properties from  lattice energy levels using chiral effective field theory}

\ShortTitle{Lattice energy levels and chiral EFT}

\author{\speaker{Zhi-Hui Guo}\\
        Department of Physics and Hebei Advanced Thin Films Laboratory, Hebei Normal University, Shijiazhuang 050024, China\\
        E-mail: \email{zhguo@hebtu.edu.cn}}
\author{Liuming Liu\\
       Institute of Modern Physics, Chinese Academy of Sciences, Lanzhou 730000, China\\
       E-mail: \email{liuming@impcas.ac.cn}}
\author{Ulf-G. Mei{\ss}ner\\
       Helmholtz-Institut f\"ur Strahlen- und Kernphysik and Bethe Center for Theoretical Physics, Universit\"at Bonn, D--53115
Bonn, Germany\\
Institute for Advanced Simulation, Institut f{\"u}r Kernphysik and J\"ulich Center for Hadron Physics, Forschungszentrum  J{\"u}lich, D-52425 J{\"u}lich, Germany \\ 
 Ivane Javakhishvili Tbilisi State University, 0186 Tbilisi, Georgia \\
       E-mail: \email{meissner@hiskp.uni-bonn.de}}       
\author{J.~A.~Oller\\
      Departamento de F\'{\i}sica, Universidad de Murcia, E-30071 Murcia, Spain\\
       E-mail: \email{oller@um.es}}    
\author{A.~Rusetsky\\
       Helmholtz-Institut f\"ur Strahlen- und Kernphysik and Bethe Center for Theoretical Physics, Universit\"at Bonn, D--53115
Bonn, Germany\\
       E-mail: \email{rusetsky@hiskp.uni-bonn.de}}    

\abstract{ We use the chiral effective field theory to study the lattice finite-volume energy levels from the meson-meson scattering. The hadron resonance properties and the scattering amplitudes at physical masses are determined from the lattice energy levels calculated at unphysically large pion masses. The results from the $\pi\eta, K\bar{K}$ and $\pi\eta'$ coupled-channel scattering and the $a_0(980)$ resonance are explicitly given as a concrete example. }

\FullConference{The 36th Annual International Symposium on Lattice Field Theory - LATTICE2018\\
		22-28 July, 2018\\
		Michigan State University, East Lansing, Michigan, USA.}

\begin{document}

\section{Introduction}

Recently, remarkable progresses on the study of meson-meson scattering have been achieved in the lattice calculation. A large amount of the finite-volume energy levels are obtained by introducing various interpolating operators and many moving frames. The rich lattice spectra encode valuable information of the scattering processes in question. For the meson-meson scattering processes that are still lacking any direct experimental measurement, the lattice spectra are specially useful. Nevertheless, most of the available lattice energy levels are calculated at unphysically large pion masses. Therefore the chiral extrapolation is generally needed at present in order to obtain physical observables from those lattice energy levels. The method with flexible algebraic parameterizations of the $K$ matrix popularly used in many lattice analyses~\cite{Moir:2016srx,Dudek:2016cru} is   efficient to perform the fits to the precise lattice energy levels, which however may not provide reliable formulas for the chiral extrapolation. In Refs.~\cite{Guo:2016zep,Guo:2018}, we have proposed to use the unitarized chiral amplitudes by including the finite-volume effects to fit the lattice energy levels, which turn out to be successful to reproduce the lattice data at unphysically large pion masses. An important advantage is that we can reliably perform the chiral extrapolation using the chiral amplitudes, which enables us to predict the physical scattering amplitudes directly from the lattice energy levels at unphysically large pion masses. In this paper we review the theoretical setups and highlight the important findings of Refs.~\cite{Guo:2016zep,Guo:2018}.

\section{Theoretical formalism}\label{sec.theo}

The basic formula in our study is the unitarized chiral amplitude given by
\begin{eqnarray} \label{eq.defut}
 T(s) = \big[ 1 + N(s)\cdot G(s) \big]^{-1}\cdot N(s)\,,
\end{eqnarray}
which is an algebraic approximation of the $N/D$ method~\cite{Oller:1998zr,Oller:2000fj}. By definition $G(s)$ only includes the $s$-channel unitarity/right-hand cuts and $N(s)$ incorporates the crossed-channel contributions. In this way, the $s$-channel unitarity is exactly implemented and the crossed-channel effects are perturbatively included order by order through the $N(s)$ function. One way to give the $G(s)$ function is 
\begin{eqnarray}\label{eq.defg}
G(s)=i\int\frac{{\rm d}^4q}{(2\pi)^4}
\frac{1}{(q^2-m_{1}^2+i\epsilon)[(P-q)^2-m_{2}^2+i\epsilon ]}\ ,\qquad
s\equiv P^2\ \,.
\end{eqnarray}
The explicit form of the $G(s)$ function, calculated with the dimensional regularization by replacing the divergence by a constant, reads~\cite{Oller:1998zr} 
\begin{eqnarray}\label{eq.gfunc}
G(s)^{\rm DR} \al=\al\frac{1}{16\pi^2}\bigg\{{a}(\mu)+\ln\frac{m_{1}^2}{\mu^2}
+\frac{s-m_{1}^2+m_{2}^2}{2s}\ln\frac{m_{2}^2}{m_{1}^2}\nonumber\\
\al\al+\frac{\sigma}{2s}\big[\ln(s-m_{2}^2+m_{1}^2+\sigma)-\ln(-s+m_{2}^2-m_{1}^2+\sigma)\nonumber\\
\al\al+\ln(s+m_{2}^2-m_{1}^2+\sigma)-\ln(-s-m_{2}^2+m_{1}^2+\sigma)\big]\bigg\}\ ,
\end{eqnarray}
with $\sigma=\sqrt{\lambda(s,m_1^2,m_2^2)}$ and $\mu$ the regularization scale. 
While for the $N(s)$ function in Eq.~\eqref{eq.defut}, it can be given by the perturbative chiral amplitudes. If only the tree-level results are considered, $N(s)$ simply equals to the partial-wave chiral amplitudes~\cite{Guo:2018,Guo:2016zep}. When the chiral loops are taken into account, one should remove the contributions from the unitarity cuts, see Refs.~\cite{Guo:2011pa,Guo:2012yt}. For the coupled-channel case, the $N(s)$ and $G(s)$ should be understood as matrices, spanned in the channel space. 

One can include the finite-volume corrections in the unitarized chiral amplitude through the $G(s)$ function. After the integration of the zeroth component $q^0$, the four-momentum integral of Eq.~\eqref{eq.defg} reduces to 
\begin{eqnarray}\label{eq.defg3d} 
 G(s)^{\rm cutoff}= \int^{|\vec{q}|<q_{\rm max}} \frac{{\rm d}^3 \vec{q}}{(2\pi)^3} \, I(|\vec{q}|) \,,
\end{eqnarray}
where 
\begin{eqnarray}
I(|\vec{q}|) = \frac{w_1+w_2}{2w_1 w_2 \,[E^2-(w_1+w_2)^2]}\,, \quad
w_i =\sqrt{|\vec{q}|^2+m_i^2} \,,  \quad E=\sqrt{s} \,. 
\end{eqnarray} 
When performing the integration of $q^0$ to obtain Eq.~\eqref{eq.defg3d}, it is convenient to work in the center of mass (CM) frame, by taking the total four-momentum $P^\mu$ as $(P^0,\vec{P}=0)$. In the following, we shall denote the quantity defined in the CM frame with an asterisk. Notice that a three-momentum cutoff $q_{\rm max}$ is introduced in Eq.~\eqref{eq.defg3d} to regularize the divergent integral. 

The finite-volume correction can be introduced by discretizing the integral of Eq.~\eqref{eq.defg3d}. By imposing the periodical boundary conditions to the cubic box with length $L$, one can replace the three-momentum integral in Eq.~\eqref{eq.defg3d} with the discrete sum~\cite{Doring:2011vk} 
\begin{eqnarray}\label{eq.gtilde}
 \widetilde{G}= \frac{1}{L^3} \sum_{\vec{n}}^{|\vec{q}^{*}|<q_{\rm max}} I(|\vec{q}^{*}|)\,,\quad 
 \vec{q}^{*}= \frac{2\pi}{L} \vec{n}, \quad  \vec{n} \in \mathbb{Z}^3 \,.
\end{eqnarray}
For a quantity defined in the finite volume, we put a tilde on top of it. 
Then the finite-volume correction $\Delta G$ to the $G(s)$ function in the infinite volume takes the form  
\begin{eqnarray}\label{eq.deltag}
\Delta G &=& \widetilde{G} - G^{\rm cutoff} \nonumber \\
 & = & \frac{1}{L^3} \sum_{\vec{n}}^{|\vec{q}^{*}|<q_{\rm max}} I(|\vec{q}^{*}|) -   \int^{|\vec{q}^{*}|<q_{\rm max}} 
\frac{{\rm d}^3 \vec{q}^{*}}{(2\pi)^3} I(|\vec{q}^{*}|)\,. 
\end{eqnarray}
We have explicitly shown that the three-momentum cutoff $q_{\rm max}$ dependence of the $\Delta G$ is weak in Ref.~\cite{Guo:2016zep}. In the finite-volume analysis, the corresponding expression of the function $G(s)$ is then given by  
\begin{eqnarray}\label{eq.gfuncfvdr}
\widetilde{G}^{\rm DR}= G^{\rm DR} + \Delta G \,.
\end{eqnarray}

In the finite box, the Lorentz invariance does not hold any more and one needs to 
explicitly calculate the corresponding expressions in different moving frames~\cite{Doring:2012eu,Gockeler:2012yj,Roca:2012rx,Fu:2011xz,Rummukainen:1995vs}. 
For the two-body scattering with the total four-momentum $P^\mu=(P^0,\vec{P})$, 
let's denote $\vec{q_1}$ and $\vec{q}_2$ as the three-momentum of the two particles, 
with $\vec{q}_1+ \vec{q}_2=\vec{P}$. The energy of the two-body system 
in the CM frame is $E=\sqrt{s}=\sqrt{(P^0)^2-|\vec{P}|^2|}$. Via the Lorentz boost 
from the moving frame to the CM frame, one can obtain the relations between 
$\vec{q}_i$ and $\vec{q}^{*}_i$
\begin{eqnarray}\label{eq.lrtr2}
 \vec{q}_i^{\,*}= \vec{q}_i + \bigg[ \bigg(\frac{E}{P^0} - 1\bigg)
\frac{\vec{q}_i\cdot \vec{P}}{|\vec{P}|^2} 
- \frac{q_i^{*\,0}}{P^0} \bigg]\vec{P}\,, 
\end{eqnarray} 
with 
\begin{eqnarray}
 q_1^{\,*\,0}= \frac{E^2+m_1^2-m_2^2}{2E} \,, \quad  q_2^{\,*\,0}= \frac{E^2+m_2^2-m_1^2}{2E} \,.
\end{eqnarray}
The corresponding formula of the discrete sum~\eqref{eq.gtilde} in the moving 
frame is then given by~\cite{Doring:2012eu}
\begin{eqnarray}\label{eq.gfuncfvmv}
\widetilde{G}^{\rm MV}= \frac{E}{P^0 L^3}  \sum_{\vec{q}}^{|\vec{q}^{\,*}|<q_{\rm max}} I(|\vec{q}^{\,*}(\vec{q})|)\,, 
\quad \vec{q}=\frac{2\pi}{L}\vec{n}\,, \quad \vec{P}=\frac{2\pi}{L}\vec{N}\,,  \quad  (\vec{n}, \vec{N}) \in \mathbb{Z}^3\,.
\end{eqnarray} 
The expression enters in the moving frame for the finite-volume analysis reads 
\begin{eqnarray}\label{eq.gfuncfvdrmv}
 \widetilde{G}^{\rm DR, MV}= G^{\rm DR} + \Delta G^{\rm MV}  \,, \quad \Delta G^{\rm MV} =   \widetilde{G}^{\rm MV} - G^{\rm cutoff}\,.
\end{eqnarray}

For the case of the $S$-wave meson-meson scattering, the finite-volume energy levels correspond to the solutions 
\begin{equation}\label{eq.detcm}
 \det[I+ N(s)\cdot \widetilde{G}^{\rm DR}] = 0\,,
\end{equation}
and 
\begin{equation}\label{eq.detmv} 
 \det[I+ N(s)\cdot\widetilde{G}^{\rm DR,MV}] = 0\,,
\end{equation}
for the CM and moving frames, respectively. 
When including higher partial waves, there will be mixing between different partial waves and the mixing patterns vary in different irreducible representations and different moving frames. We refer to Ref.~\cite{Guo:2018} and references therein for details. 

\section{Results}

\begin{figure}[htbp]
 \centering
\includegraphics[width=0.6\textwidth,angle=-0]{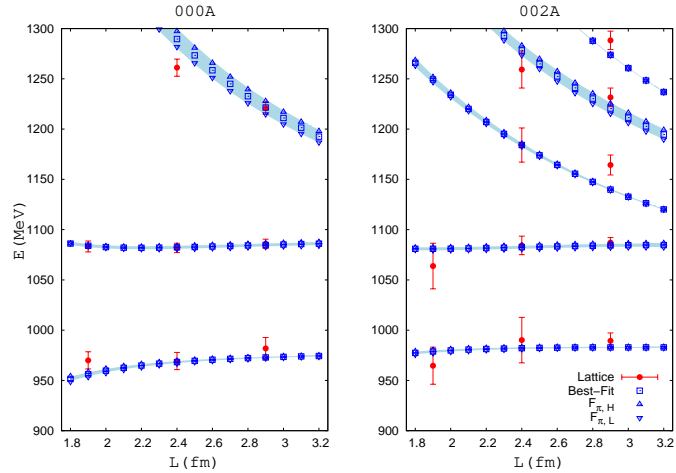} 
\caption{ Reproduction of the lattice energy levels, which are taken from Ref.~\cite{Dudek:2016cru}, for the cases of $000A$ and $002A$ at LO. The pion mass used in the lattice calculation is $m_\pi=391$~MeV~\cite{Dudek:2016cru}. The squares correspond to the central-value results from the best fit by taking $F_\pi=105.9$~MeV, with the surrounding shaded areas the statistical uncertainties at one-sigma level. 
The upwards ($F_{\pi, {\rm H} }$) and downwards ($F_{\pi, {\rm L}}$) triangles, are the results with the upper and lower limits of $F_\pi=105.9 \pm 3.6~\,{\rm MeV}$, respectively. }
   \label{fig.lolev000}
\end{figure} 

\begin{figure}[htbp]
   \centering
   \includegraphics[width=0.6\textwidth,angle=-0]{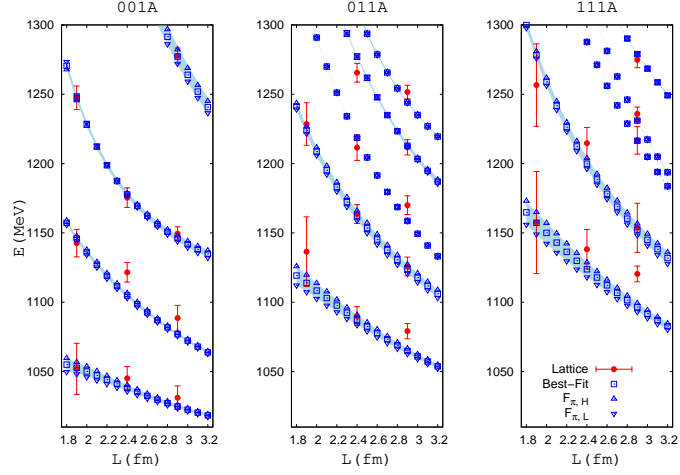} 
  \caption{Reproduction of the lattice energy levels from Ref.~\cite{Dudek:2016cru} for the cases of $001A$, $011A$ and $111A$ at leading order. See Fig.~1 for the meaning of notations.   }
   \label{fig.lolev001}
\end{figure}

\begin{figure}[htbp]
   \centering
   \includegraphics[width=0.99\textwidth,angle=-0]{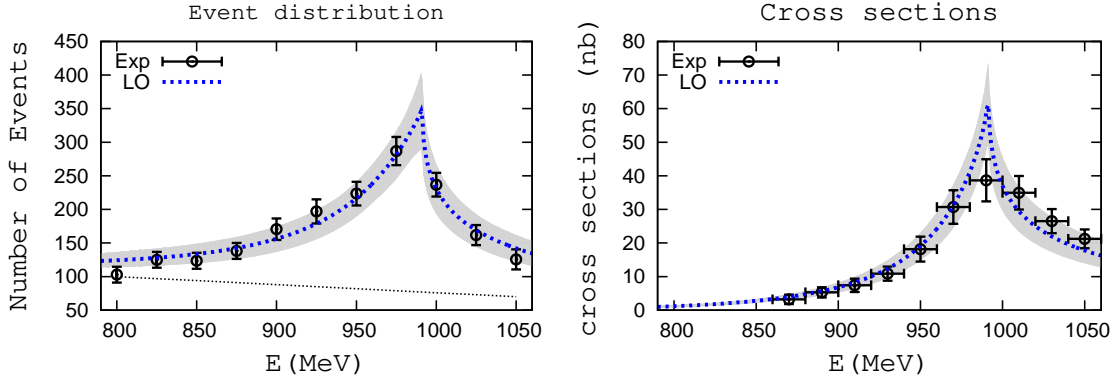} 
  \caption{Reproduction of the $\pi\eta$ event distributions from Ref.~\cite{Armstrong:1991rg} (left panel) and the cross sections of $\gamma\gamma\to \pi\eta$ from Ref.~\cite{Uehara:2009cf}. The black dotted line in the left panel is the background from Ref.~\cite{Armstrong:1991rg}. The blue dotted lines in both panels stand for the central-value results at LO and the surrounding shaded areas are the error bands at one-sigma level. }
   \label{fig.fitexp}
\end{figure}

With the setups in Sec.~\ref{sec.theo}, we are ready to fit the lattice energy levels to determine the unknown parameters, including the chiral low energy constants in the $N(s)$ and the subtraction constants in the $G(s)^{\rm DR}$. Here we focus on the $S$-wave $\pi\eta$, $K\bar{K}$ and $\pi\eta'$ coupled-channel scattering. At leading order (LO), there are only tree-level contributions from chiral perturbation theory and the $S$-wave perturbative chiral amplitudes read~\cite{Guo:2011pa} 
\begin{eqnarray}\label{eq.pwlo}
 T_{J=0}^{I=1, \pi\eta\to\pi\eta}(s) &=& \frac{(c_\theta-\sqrt2 s_\theta)^2m_\pi^2}{3 F_\pi^2}\,,\nonumber \\
 T_{J=0}^{I=1, \pi\eta\to K\bar{K}}(s) &=& \frac{c_\theta(3m_\eta^2+8 m_K^2 + m_\pi^2-9s)+2\sqrt{2}s_\theta(2m_K^2+m_\pi^2) }{6\sqrt{6} F_\pi^2}\,,\nonumber \\
  T_{J=0}^{I=1, \pi\eta\to\pi\eta'}(s) &=& \frac{(\sqrt{2}c_\theta^2-c_\theta s_\theta - \sqrt2 s_\theta^2)m_\pi^2}{3 F_\pi^2}\,,\nonumber \\
 T_{J=0}^{I=1, K\bar{K}\to K\bar{K}}(s) &=& \frac{s}{4 F_\pi^2}\,,\nonumber \\
 T_{J=0}^{I=1, K\bar{K}\to \pi\eta'}(s) &=& \frac{s_\theta(3m_{\eta'}^2+8 m_K^2 + m_\pi^2-9s) - 2\sqrt{2}c_\theta(2m_K^2+m_\pi^2) }{6\sqrt{6} F_\pi^2}\,,\nonumber \\
 T_{J=0}^{I=1, \pi\eta'\to\pi\eta'}(s) &=& \frac{(\sqrt2 c_\theta + s_\theta)^2m_\pi^2}{3 F_\pi^2}\,,
\end{eqnarray}
with $I$ short for isospin, $c_\theta=\cos\theta$, $s_\theta=\sin\theta$ and $\theta$ the LO mixing angle of $\eta$-$\eta'$. 
If only including the LO results, the matrix elements of $N(s)$ in Eq.~\eqref{eq.defut} simply equal to the corresponding expressions in Eq.~\eqref{eq.pwlo}.  At LO, the reproduction of the lattice data~\cite{Dudek:2016cru} can be seen Figs.~\ref{fig.lolev000} and \ref{fig.lolev001}. In the meanwhile, we also 
include two types of experimental data in the fits, including the $\pi\eta$ invariant mass distributions from Ref~\cite{Armstrong:1991rg} and the $\gamma\gamma\to\pi\eta$ cross sections from Ref.~\cite{Uehara:2009cf}. In order to fit the invariant mass distributions and cross sections, we need to introduce several additional parameters~\cite{Guo:2016zep}, which are irrelevant to the $\pi\eta$, $K\bar{K}$ and $\pi\eta'$ scattering. The fit results of the invariant mass distribution and cross sections can be seen in Fig.~\ref{fig.fitexp}.

\begin{figure}[htbp]
   \centering
   \includegraphics[width=0.99\textwidth,angle=-0]{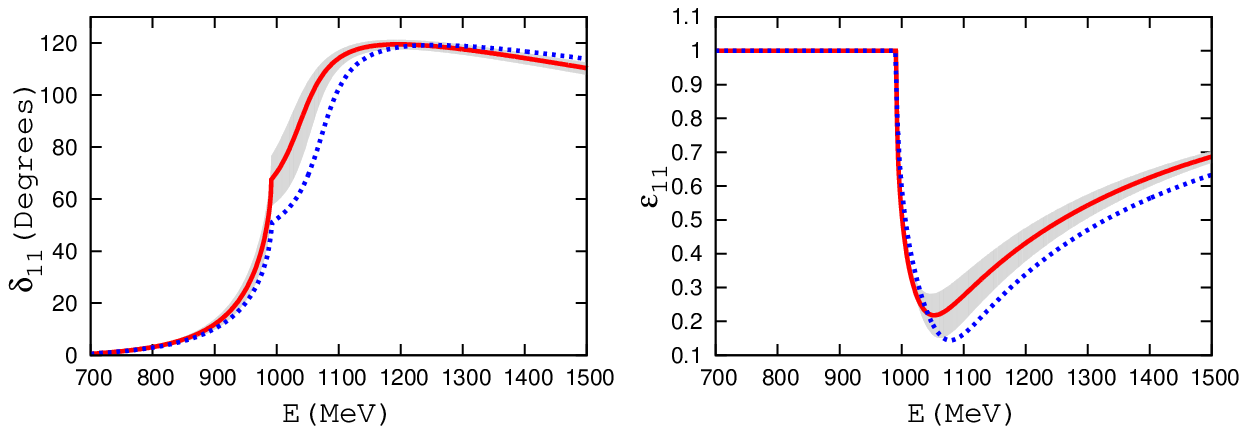} 
  \caption{ Prediction of the phase shifts (left panel) and inelasticities (right panel) from the $\pi\eta \to \pi\eta$ scattering obtained at physical masses at LO. The red solid lines are the central-value results from the LO fit with a universal pion decay constant in all the scattering amplitudes. The shadow areas stand for the statistical uncertainties at one-sigma level. The blue dotted lines correspond to the curves by using $F_\pi$ and $F_K$ in the corresponding amplitudes~\cite{Guo:2016zep}. } 
   \label{fig.lophasephy}
\end{figure}

After the determination of the unknown parameters from the fits, we can perform the chiral extrapolation to obtain the scattering amplitudes at physical meson masses. We give the predictions of the phase shifts and inelasticities of the $\pi\eta\to\pi\eta$ scattering in Fig.~\ref{fig.lophasephy}. We find that there is a resonance pole on the fourth Riemann sheet that can be identified as the $a_0(980)$, which is $(1037_{-14}^{+17}-i44_{-9}^{+6})$~MeV. The corresponding residue is $3.8_{-0.2}^{+0.3}$~GeV for the $\pi\eta$ channel and the ratios of the $K\bar{K}$ and $\pi\eta'$ channels to the $\pi\eta$ channel are $1.43_{-0.03}^{+0.03}$ and $0.05_{-0.01}^{+0.01}$, respectively.  

In this paper, we have reviewed the recent works in Refs.~\cite{Guo:2016zep,Guo:2018} to demonstrate that the chiral amplitudes with the inclusion of the finite-volume effects are useful to extract physical scattering observables and resonance properties from the discrete lattice energy levels obtained at unphysically large meson masses. This framework can be straightforwardly applied to other scattering processes, and it can provide valuable physical quantities that are not available from the experimental measurements~\cite{Guo:2018}.

\section*{Acknowledgements}
This work is funded in part by the NSFC under grants No.~11575052, the Natural Science Foundation of Hebei Province under contract No.~A2015205205. This work is also partially supported by the Sino-German Collaborative Research Center ``Symmetries and the Emergence of Structure in QCD'' (CRC~110) co-funded by the DFG and the NSFC. JAO would like to thank partial financial support to the MINECO (Spain) and EU grant FPA2016-77313-P. UGM and AR acknowledge the support from Volkswagenstiftung under contract No. 93562.  AR acknowledges the support from Shota Rustaveli National Science Foundation (SRNSF), grant No. DI-2016-26. The work of UGM was also supported by the Chinese Academy of Sciences (CAS) President's International Fellowship Initiative (PIFI) (Grant No. 2018DM0034). LL acknowledges the support from the Key Research Program of the Chinese Academy of Sciences, Grant NO. XCPB09.


\begin{thebibliography}{99}
  
  

\bibitem{Moir:2016srx} 
  G.~Moir, M.~Peardon, S.~M.~Ryan, C.~E.~Thomas and D.~J.~Wilson,
  JHEP {\bf 1610}, 011 (2016).

\bibitem{Dudek:2016cru} 
  J.~J.~Dudek {\it et al.} [Hadron Spectrum Collaboration],
  Phys.\ Rev.\ {\bf D93}, no. 9, 094506 (2016).
  
  
\bibitem{Guo:2016zep} 
  Z.~H.~Guo, L.~Liu, U.-G.~Mei{\ss}ner, J.~A.~Oller and A.~Rusetsky,
  Phys.\ Rev.\ D {\bf 95}, no. 5, 054004 (2017).
  
\bibitem{Guo:2018} 
  Z.~H.~Guo, L.~Liu, U.-G.~Mei{\ss}ner, J.~A.~Oller and A.~Rusetsky, in preparation. 

 
  
\bibitem{Oller:1998zr} 
  J.~A.~Oller and E.~Oset,
  Phys.\ Rev.\ D {\bf 60}, 074023 (1999).

\bibitem{Oller:2000fj} 
  J.~A.~Oller and U.-G.~Mei{\ss}ner,
  Phys.\ Lett.\ B {\bf 500}, 263 (2001).  
 
  
  
 
  
 
 
  
%
%
    
\bibitem{Guo:2011pa} 
  Z.~H.~Guo and J.~A.~Oller,
  Phys.\ Rev.\ {\bf D84}, 034005 (2011).
  
  
\bibitem{Guo:2012yt} 
  Z.~H.~Guo, J.~A.~Oller and J.~Ruiz de Elvira,
  Phys.\ Rev.\ D {\bf 86}, 054006 (2012).
 
\bibitem{Doring:2011vk} 
  M.~D\"oring, U.-G.~Mei{\ss}ner, E.~Oset and A.~Rusetsky,
  Eur.\ Phys.\ J.\ A {\bf 47}, 139 (2011).
  
\bibitem{Doring:2012eu} 
  M.~D\"oring, U.-G.~Mei{\ss}ner, E.~Oset and A.~Rusetsky,
  Eur.\ Phys.\ J.\ A {\bf 48}, 114 (2012).
  
  
\bibitem{Gockeler:2012yj} 
  M.~G\"ockeler, R.~Horsley, M.~Lage, U.-G.~Mei{\ss}ner, P.~E.~L.~Rakow, A.~Rusetsky, 
  G.~Schierholz and J.~M.~Zanotti,
  Phys.\ Rev.\ D {\bf 86}, 094513 (2012).
%
  
  
\bibitem{Roca:2012rx} 
  L.~Roca and E.~Oset,
  Phys.\ Rev.\ D {\bf 85}, 054507 (2012).
  
  
\bibitem{Fu:2011xz} 
  Z.~Fu,
  Phys.\ Rev.\ D {\bf 85}, 014506 (2012).
  
  
\bibitem{Rummukainen:1995vs} 
  K.~Rummukainen and S.~A.~Gottlieb,
  Nucl.\ Phys.\ B {\bf 450}, 397 (1995).
  
  
\bibitem{Armstrong:1991rg} 
  T.~A.~Armstrong {\it et al.} [WA76 and Athens-Bari-Birmingham-CERN-College de France Collaborations],
  Z.\ Phys.\ C {\bf 52}, 389 (1991).
  
\bibitem{Uehara:2009cf} 
  S.~Uehara {\it et al.} [Belle Collaboration],
  Phys.\ Rev.\ {\bf D80}, 032001 (2009).
  
\end{thebibliography}
\end{document}